\documentclass[a4paper]{jpconf}
\usepackage{graphicx}
\usepackage{amsmath,amssymb,amsfonts,amsthm}

\newcommand{\be}{\begin{equation}}
\newcommand{\ee}{\end{equation}}
\newcommand{\ben}{\begin{equation*}}
\newcommand{\een}{\end{equation*}}
\newcommand{\bea}{\begin{eqnarray}}
\newcommand{\eea}{\end{eqnarray}}
\newcommand{\bean}{\begin{eqnarray*}}
\newcommand{\eean}{\end{eqnarray*}}
\newcommand{\brr}{\begin{array}}
\newcommand{\err}{\end{array}}
\newcommand{\bc}{\begin{center}}
\newcommand{\ec}{\end{center}}

\newcommand{\vev}[1]{\mbox{$\langle #1 \rangle $}}
\renewcommand{\lsim}{\,\raisebox{-0.6ex}{$\buildrel < \over \sim$}\,}

\newcommand{\bk}{{\mathbf k}}

\newcommand{\by}{{\mathbf y}}
\newcommand{\bx}{{\mathbf x}}

\newcommand{\HH}{{\cal H}}
\newcommand{\MM}{{\cal M}}

\newcommand{\al}{\alpha}

\newcommand{\de}{\delta}
\newcommand{\De}{\Delta}
\newcommand{\ep}{\epsilon}
\newcommand{\ga}{\gamma}

\newcommand{\Om}{\Omega}
\newcommand{\om}{\omega}

\newcommand{\dd}{\partial}
\newcommand{\gsim}{\,\raisebox{-0.6ex}{$\buildrel > \over \sim$}\,}

\begin{document}
\title{Gravitational waves from cosmological phase transitions}

\author{Ruth Durrer}
    
\address{D\'epartement de Physique  Th\'eorique,
  Universit\'e de Gen\`eve\\ 24 Quai E. Ansermet, 1211 Gen\'eve 4,
  Switzerland}

\ead{ruth.durrer@unige.ch}

\begin{abstract}
In this talk I discuss the generation of a stochastic background of 
gravitational waves during a first order phase transition. I present
simple general arguments which explain the main features of the
gravitational wave spectrum like the $k^3$ power law growth on large scales
and a estimate for the peak amplitude. In the second part I concentrate
on the electroweak phase transition and argue that the nucleosynthesis bound 
on its gravitational wave background seriously limits seed magnetic fields which
may have been generated during this transition.  
\end{abstract}

\section{Introduction}
 The Universe has expanded and cooled down from a very hot
  initial state to (presently) $2.7^o$K. It seems likely that it underwent
 several phase transitions during its evolution of adiabatic expansion.
It has already been proposed in the 80ties, that a cosmological phase 
transition can lead to the generation of a stochastic gravitational wave
background~\cite{mag1}.
If the phase transitions are of second order or only a crossover
as it has been obtained from standard model calculations, we do not 
expect them to lead to an observable cosmological signal. However,
if a phase transition is first order, it proceeds via bubble nucleation
which is a highly inhomogeneous process. Some of the kinetic energy
of the rapidly expanding bubbles is transferred to the cosmic
plasma and leads to turbulence. If small seed magnetic fields are present,
they are amplified by the turbulent motion which becomes MHD
turbulence. Since the fluid and magnetic Reynolds numbers in the hot 
cosmic plasma are extremely high, MHD turbulence persists even
after the phase transition, where it freely decays. These inhomogeneities
int the energy and momentum of the cosmic plasma generate a stochastic 
background of gravitational waves. Here we want to investigate this 
background. 

\subsection{Events in the early Universe}
The following important events have most probably taken place in the early 
Universe and they may, under certain circumstances have induced a 
background of stochastic gravitational waves.
 \begin{itemize}
\item  {\em Inflation} is supposed to have generated the scalar fluctuations in the geometry
and energy density which have led to the observed anisotropies in the cosmic
microwave background (CMB)~\cite{Muk,myBook}.
Simple inflationary models do not only induce scalar perturbations but the quantum fluctuations
in the gravitational spin-2 degrees of freedom also lead to a scale invariant background
 of gravitational waves of similar amplitude. It is one of the most important goals of 
 present CMB experiments like
 Planck~\cite{Planck} to observe the B-polarization which this gravitational wave background 
 should induce in the CMB photons.
 \item {\em Pre-heating:} At the end of inflation the inflaton decays into other degrees
 of freedom which eventually reheat the Universe and lead to a thermal bath of relativistic degrees of freedom containing especially also the standard model particles. The details of this process 
are complicated and in some cases they generate inhomogeneities leading to 
 a  significant stochastic background of gravitational waves~\cite{preheat}. Using the 
standard relation between temperature and time in a Friedmann Universe filled 
with relativistic particles~\cite{myBook}, we obtain for a reheat
 temperature $T_i$
 \bea
 T_i  &\simeq& 10^{14}{\rm GeV} \,, \quad t_i = 2.3{\rm sec}  \left(\frac{1{\rm MeV}}{ T_i}\right)^2 
 g_{\rm eff}(T)^{-1/2} \simeq 10^{-32}{\rm sec}\, , \\
  \eta_i &=& \frac{1}{a(t_i)H(t_i)} =2t_i(1+z_i) \simeq 10^{-7} {\rm sec,}\quad
 \om_i \gsim  \frac{1}{\eta_i } \simeq  10^{7}{\rm Hz,} \label{e:omi}
 \eea
 $ g_{\rm eff}(T)$ denotes the effective number of relativistic degrees of freedom at
 temperature $T$~\cite{myBook}.
To find the conformal time $\eta_i$ we have used that in a radiation 
dominated universe 
$$\eta= \frac{1}{\HH}=\frac{1}{aH} =\frac{1+z}{H} =2t(1+z)\,,$$
where $\HH =aH$ is the comoving Hubble rate and $a$ is the scale factor 
normalized to one today, $a(t_0)=1$.
Furthermore, since $ g_{\rm eff}(aT)^3$ is constant during adiabatic expansion
$$ 1+z =\frac{1}{a}= \frac{ g_{\rm eff}^{1/3}T}{g_0^{1/3}T_0} = 
\frac{10^{10}}{2.35}\left(\frac{T}{1{\rm MeV}}\right)
   \left(\frac{g_{\rm eff}}{2}\right)^{1/3} \,,$$
so that
\be
  \eta= 0.67\times 10^{10}\left(\frac{T}{1{\rm MeV}}\right) g_{\rm eff}^{1/3}t
 \simeq 1.5\times 10^{10}{\rm sec} \left(\frac{1{\rm MeV}}{T}\right)g_{\rm eff}(T)^{-1/6}\,.
\ee
We have set $k_B=1$ and measure temperature in MeV. In the following
we use units with $c=\hbar=1$ so that length and time have the same 
units as inverse energy.

 For the $\gsim$ sign in eq.~(\ref{e:omi}) we have used that the typical 
frequency generated by such an event must be larger than the inverse of the 
correlation scale which is always smaller than the horizon size at the given 
epoch. Note that $\om_i$ denotes the comoving frequency and we have normalized 
the scale factor to unity today, so that co-moving frequencies, length scales 
and times correspond to physical scales today. 

So far no experiment which can detect gravitational waves with frequencies 
as high as $\om_i$ has been proposed.
 \item {\em The electroweak transition:} In the standard model, for realistic
values of the Higgs mass, the electroweak transition is not even second order, but only a 
crossover~\cite{ew}.  However, small deviations from the standard model e.g. 
in the Higgs sector or supersymmetric models~\cite{devew} can can lead to a 
first order electroweak phase transition at temperature
 \bea
 T_{ew} \simeq 10^{2}{\rm GeV ,}&&\quad t_{ew} \simeq 10^{-10}{\rm sec ,} \\
   \eta_{ew} \simeq  10^5{\rm sec ,}&& \quad  \om_{ew} \gsim 10^{-5} {\rm Hz. }
  \eea 
  If the electroweak phase transition is strongly first order, the bubbles 
expand very rapidly
  and the duration of the phase transition is a small fraction, maybe 1\% of the Hubble time.
  In this case one expects a peak frequency of  $\om_{ew} \sim 100/\eta_{ew} \simeq 10^{-3}$Hz.
  This is the frequency of best sensitivity for the gravitational wave 
satellite LISA proposed for launch in 2018~\cite{LISA}. 
 \item {\em Confinement transition:} Also in this case, standard lattice QCD 
 calculations~\cite{QCD} predict a simple crossover. However, the 
 confinement transition can be first order if the neutrino chemical potentials 
are sufficiently large (still within the limits allowed by 
nucleosynthesis)~\cite{DS}. Models which lead to such high neutrino chemical  
potentials have been proposed for leptogenesis with sterile 
neutrini~\cite{nuchem}. The temperature of the QCD transition is about
 \bea
 T_{c} &\simeq& 10^{2}{\rm MeV ,} \quad  t_{c} \simeq 10^{-5}{\rm sec ,}\\ 
  \eta_{c} &\simeq&  10^7{\rm sec, } ~\quad ~\om_c \gsim 10^{-7} {\rm Hz .}
 \eea
Gravitational waves in this frequency range could be observed by the pulsar 
timing array which has been proposed recently~\cite{pulsar}.
 \item {\em Surprises:} Last but not least we may hope to find a gravitational 
wave background from a phase transition at some other not predicted 
temperature which then would indicate new physics at the corresponding 
energy scale, see, e.g~\cite{surprise}. For example, a first order phase 
transition at a critical 
temperature $T_* \simeq 10^7$GeV may lead to a gravitational wave background 
with a peak frequency of about 100Hz which is the best sensitivity of LIGO. 
As we shall argue below, in a very optimistic scenario it is even conceivable 
that such a background might be detected  by the advanced configuration 
of the LIGO~\cite{LIGO} experiment.
 \end{itemize}
A more comprehensive overview on first order phase transitions as sources for 
gravitational waves can be found in Ref.~\cite{over}.

In the following we present a semi-analytical determination of the gravitational
wave (GW) signal from a first order phase transition in terms of a few free 
parameters. We then apply our general results to the electroweak phase 
transition and discuss some consequences. 
 
\section{General results}
Before we outline their derivation, let us present the main results. We consider a
phase transition at $T=T_*$ which generates (relativistic) anisotropic 
stresses which source gravitational waves. We assume that the energy density 
of this source is $\rho_X$. Since the source is relativistic its anisotropic 
stresses are of the same order of magnitude. The GW spectrum then has the 
following main properties:
 \begin{itemize}
\item  {\em The peak frequency} (the inverse of the correlation scale) is 
larger than the Hubble rate, $k_* \gsim \HH_*$. Here we always compare 
comoving quantities which become physical scales today.
 \item The  GW {\em energy density} is of the order of 
   \bea 
\Om_{gw}(\eta_0) &\sim& \ep\; \Om_{\rm rad}(\eta_0)\left(\frac{\Om_X(\eta_*)}{
\Om_{\rm rad}(\eta_*)}\right)^2 \qquad \mbox{where} \nonumber \\  
\label{e:Omgwsimp} 
\ep &=&  \left\{\begin{array}{lll} \left(\HH_*\De\eta_*\right)^2 & \mbox{if }&  
\HH_*\De\eta_* < 1 \\
   1 &  \mbox{if }&  \HH_*\De\eta_* \ge1 \, . \end{array} \right. 
   \eea
Here  $\De\eta_*$ is the duration over which the source is active, typically 
the duration of the phase transition. $\Om$ is the density parameter, 
$\Om_X =\rho_X/\rho_c$ where $\rho_X$ is the energy density of the 
component $X$ and $\rho_c$ is the critical density.
 \item On large scales, $k\ll k_*$ the spectrum is blue, 
    $$ \frac{d\Om_{gw}(k) }{d\log(k)} \propto k^3 ~ , \qquad \Om_{gw}  = 
       \int \frac{dk}{k} \frac{d\Om_{gw}(k) }{d\log(k)}\,. $$
The second equality defines the the gravitational wave energy per logarithmic 
frequency interval, $\frac{d\Om_{gw}(k) }{d\log(k)}$.
 \item  On small scales, $k\gg k_*$ the spectrum decays. The decay law depends 
on the details of the source. If the anisotropic stress spectrum of the source
decays with power $k^{-\ga}$, and the source is totally coherent in time
(see below), the gravitational wave energy density decays like $k^{-(\ga+1)}$.
If the source is coherent for about one wavelength,  the gravitational wave 
energy density decays like $k^{-(\ga-1)}$ and if the source is incoherent, the 
gravitational wave energy density decays like $k^{-(\ga-3)}$. see 
Ref.~\cite{geraldine2}. The unequal time correlator of the source is discussed 
in the next section. 
 \end{itemize}

\section{Generation of gravitational waves}

Gravitational waves are sourced by fluctuations in the energy momentum tensor 
which have a non-vanishing spin-2 contribution. The perturbed metric is of 
the form
\be ds^2 = a^2\left(-d\eta^2 + (\ga_{ij} + 2 h_{ij})dx^idx^j\right) \,, \ee
where $\ga_{ij}$ is the metric of a 3--space of constant curvature $K$
and $h_{ij}$ is transverse and traceless. In Fourier space $k^ih_{ij} =h^i_i =0$. 

Einstein's equation to first order in $h_{ij}$ gives (we set $K=0$ in what follows) 
\be \left(\dd^2_\eta +2\HH\dd_\eta +k^2\right)h_{ij} = 8\pi G a^2\Pi_{ij} \,. \ee
Here $\Pi_{ij}(\bk,\eta) $ is the Fourier component of the tensor type 
(spin-2) anisotropic stress of the source and $\HH =\frac{a'}{a}$. A prime 
denotes the derivative w.r.t. conformal time $\eta$.

During a first order phase transition, anisotropic stresses can be generated by
colliding bubbles and by inhomogeneities in the cosmic fluid (e.g. turbulence, 
magnetic fields)~\cite{KKTW}. 

For a duration of the phase transition given by $\De\eta_*$, the final bubble 
size, i.e. the typical size of the bubbles when they start to coalesce and the 
phase transition terminates, is $R_*=v_b\De\eta_*$,  where $v_b$ is the bubble 
velocity. Typically  $\De\eta_*<\HH^{-1}$. As we shall see, depending on the 
unequal time correlation properties of the anisotropic stress, the peak 
frequency can be given either by the time scale, $k_* \simeq (\De\eta_*)^{-1}$,
or by the spatial scale $k_* \simeq R_*^{-1}$. For the most promising case of a 
strongly first order phase transition, the bubble velocity is however close to 
the speed of light, $v_b\simeq 1$ so that this distinction is not very relevant.

\section{The spectrum}
Because of causality, the correlator $\langle \Pi_{ij}(\eta_1,\bx) ]
\Pi_{lm}(\eta_2,\by)\rangle  = \MM_{ijlm}(\eta_1,\eta_2,\bx-\by)$ is a 
function of compact support. For distances
 $|\bx-\by|>\max(\eta_1,\eta_2)$, $\MM \equiv 0$. Therefore, the spatial 
Fourier transform, $ \MM_{ijlm}(\eta_1,\eta_2,\bk)$ is analytic 
in $\bk$ around $\bk=0$. We decompose $\Pi_{ij}$ into two helicity modes 
which we assume to be uncorrelated so that the spectrum is even under parity
    $$\Pi_{ij}(\eta,\bk) = e_{ij}^+\Pi_+(\eta,\bk) +  e_{ij}^-\Pi_-(\eta,\bk)$$ 
 $$ \langle \Pi_+(\eta,\bk) \Pi_+^*(\eta',\bk')\rangle = \langle 
\Pi_-(\eta,\bk) \Pi_-^*(\eta',\bk')\rangle 
	  =(2\pi)^3\de^3(\bk-\bk')\rho_X^2P_\Pi(\eta,\eta',k)$$
   $$\langle \Pi_{+}(\eta,\bk) \Pi_{-}^*(\eta',\bk')\rangle = 0\,.$$
Here $\rho_X$ is the energy density of the component $X$ with anisotropic 
stress $\Pi$ which has been factorized in order to keep 
$k^3P_\Pi(\eta,\eta',k)$ dimensionless. Causality implies that the function 
$P_\Pi(\eta,\eta',k)$ is analytic in $k$. We therefore expect it
to start out as white noise and to decay beyond a certain correlation scale 
   $k_*(\eta,\eta')>\min(1/\eta,1/\eta')$.   

\begin{figure}[ht]
\begin{center}
\includegraphics[width=6cm]{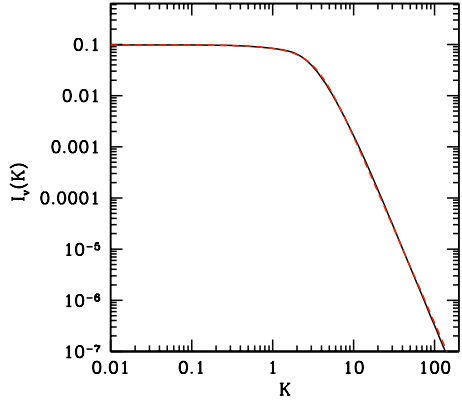}
\end{center}
\caption{\label{f:spec} The equal time anisotropic stress power 
spectrum from turbulence. On large scales the spectrum is scale invariant, 
on scales much smaller than the correlation scale it falls of  like 
$k^{-11/3}$ which is determined by the Kolmogorov turbulence assumed for the 
velocity  power spectrum. The black line is the numerical result and the 
red dashed line a simple fit. The normalization is arbitrary.
Figure from~\cite{geraldine2}.}
\end{figure}

If the gravitational wave source is active only for a short duration 
$\De\eta_*$ (less than one Hubble time), we can neglect the damping term 
$2\HH$ in the equation of motion for $h$ during the time when $\Pi$ is active. 
Here $h$ denotes either of the two helicities of the gravitational wave and 
$\Pi$ is the anisotropic stress source of the corresponding helicities. 
The solution with vanishing initial conditions is then
\bean
 h(\bk,\eta) &=& \frac{8i\pi Ga_*^3}{6ak}\left[e^{-ik\eta}
\int_{\eta_*}^{\eta_*+\De\eta_*}d\eta' e^{ik\eta'} \Pi(\eta',\bk) + \right. \\
 && \qquad \qquad \left.  e^{ik\eta}\int_{\eta_*}^{\eta_*+\De\eta_*}d\eta' 
e^{-ik\eta'} \Pi(\eta',\bk)\right] \\
 &=& \frac{8i\pi Ga_*^3}{6ak}\left[e^{-ik\eta}\Pi(k,\bk) + e^{ik\eta}\Pi(-k,\bk)
   \right]\, , \qquad \eta>\eta_*+\De\eta_*  \,.
\eean

The gravitational wave energy density is given by
$$ \rho_{gw}(\eta,\bx) = \frac{1}{32\pi G a^2}\langle \dd_\eta h_{ij}(\eta,\bx) 
    \dd_\eta h^*_{ij}(\eta,\bx)\rangle \,. $$
If the Universe is radiation dominated at the time $\eta_*$ when the 
gravitational waves are generated, this leads to the gravitational wave
energy spectrum 
\be \label{e:Omgw} 
\frac{d\Om_{gw}}{d\ln(k)}(\eta_0) =\frac{12\Om_{\rm rad}(\eta_0)}{\pi^2}
\left(\frac{\Om_{X}(\eta_*)}{\Om_{\rm rad}(\eta_*)}\right)^2
\HH_*^2k^3{\rm Re}[P_\Pi(k,k,k)]\,.
\ee
Here
\be\label{e:Ptt'k} 
P_\Pi(\om,\om',k) \equiv \int_{-\infty}^{-\infty}d\eta\int_{-\infty}^{-\infty}
d\eta' P_\Pi(\eta,\eta',k) e^{i(\om\eta-\om'\eta')} \,. 
\ee
  On large scales, $k< k_*$ the GW energy density from a 'causal' source always scales
  like $k^3$. This remains valid also for long duration sources. 
The only change for long lived sources is that the Green function $e^{ik(\eta-\eta')}$ of the 
wave operator in Minkowski spacetime has to be replaced by the Green function of the wave 
operator in an expanding universe which in the case of a radiation dominated universe is
given in terms of spherical Bessel functions.

The scale $1/k_*$ is the correlation scale which is smaller than the co-moving 
Hubble scale $1/k_*< 1/\HH_* = \eta_*$. The behavior or the spectrum close 
to the peak and its decay rate on smaller scales 
depend on the source characteristics, on its temporal behavior and 
on its power spectrum.

\begin{figure}[ht]
\begin{center}
\includegraphics[width=11cm]{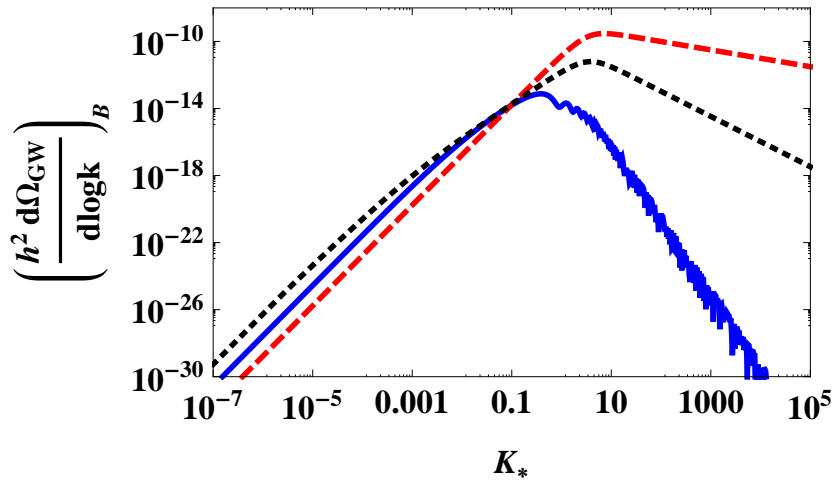}
\end{center}
\caption{\label{f:coinco}
The GW energy density spectrum for a totally incoherent source (red, long-dashed), 
a totally coherent source (blue solid) and for a source with finite coherence time 
(black, short-dashed). The parameters are: 
$T_*=100$ GeV, $\De\eta_*\mathcal{H}_*=0.01$, $\Om_X/\Om_{\rm rad}=2/9$~~ 
($\vev{v^2}=1/3$). The horizontal variable is $K_*=kR_* / (2\pi)$.
Note that the peak position is roughly at $k=\pi/\De\eta_*=$ a few, hence
$K_*\simeq R_*/\De\eta_* =v_b$ in the totally coherent case and at  
$k=\pi/R_*=$ a few, hence $K_*\simeq 1$  in the incoherent case and in 
the case with finite coherence length. In order to clearly separate the peak 
positions, $v_b \simeq 0.1$ has been chosen.
Figure from~\cite{geraldine2}.}
\end{figure}

Let us now briefly discuss the different results which have been obtained for 
the position of the peak of the gravitational wave spectrum. As we see from 
eqs.~(\ref{e:Ptt'k}) and (\ref{e:Omgw}), to determine the the gravitational wave 
energy density we must know the unequal time correlator
of the of the anisotropic stresses, $P_\Pi(\eta,\eta',k)$. This is not well known from numerical 
simulations but there exist different simple possibilities: 
\begin{itemize}
\item  A {\em totally coherent source} is one for which the time dependence is 
deterministic. In this case the randomness is solely in the initial conditions
and $\Pi(\eta,\bk) = f(\eta,k)\Pi(\eta_*,\bk)$ for some deterministic function $f$.
For the unequal time correlator $P_\Pi$ this leads to
$$P_\Pi(\eta,\eta',k)=\sqrt{P_\Pi(\eta',\eta',k)P_\Pi(\eta,\eta,k)}\,. $$
It has been shown in Refs.~\cite{thomas,geraldine2} that the peak position 
of the GW spectrum $\propto k^3P_\Pi(k,k,k)$  for a totally coherent source 
is given by the peak of the {\em temporal} Fourier transform of the source, 
i.e. $k_*=(\De\eta_*)^{-1}$.
\item For a {\em source with finite coherence time} we can model the unequal 
time correlator of the of the anisotropic stresses, by 
$$
P_\Pi(\eta,\eta',k)=\sqrt{P_\Pi(\eta',\eta',k)P_\Pi(\eta,\eta,k)}
\Theta(x_c-|\eta-\eta'|k)\, , \quad x_c \sim 1 \,.
$$ 
This ansatz assumes that the source is totally coherent if the time difference 
is smaller than about one wavelength and uncorrelated for larger time 
differences. This ansatz resembles the suggestion by Kraichnan~\cite{Kraichnan} 
for turbulence and has therefore been used for both turbulence and magnetic 
fields in Ref.~\cite{geraldine2}. However, this ansatz, like also the Kraichan 
decorrelation ansatz still lacks numerical confirmation. 

In this case, the peak of the GW spectrum is  determined by
the peak of the {\em spatial} Fourier transform of the source.
\item A {\em totally incoherent source} is uncorrelated for different times. Its  unequal 
time correlator can be modeled as
$$
P_\Pi(\eta,\eta',k)=P_\Pi(\eta',\eta',k)\De\eta_*\de(\eta-\eta') \,.
$$
Here the factor $\De\eta_*$ is present to make the dimensions right, it can be replaced 
by some other short correlation time. In this case, the GW spectrum is also determined by
the peak of the {\em spatial} Fourier transform of the source.
\end{itemize}
The resulting gravitational wave spectra from the different hypotheses for the
unequal time correlator are shown in Fig.~\ref{f:coinco}. They are derived in detail in 
Ref.~\cite{geraldine2}.

\section{The electroweak phase transition}
According to the standard model, the electroweak transition is not even second order, but 
only a cross-over. Then, this transition does not lead to the formation of gravitational waves.
Even though the Higgs vacuum expectation value does cross-over from zero to a finite 
value, the corresponding Higgs field fluctuations have a very short correlation length of
the order of the inverse temperature and do not invoke significant gravitational field fluctuations.

However, if the standard model is somewhat modified e.g. in the Higgs sector, or in certain 
regions of the MSSM parameter space, the electroweak phase transition can become first 
order, even strongly first order~\cite{devew}. Then it proceeds via the formation of large 
bubbles of true vacuum. The bubbles themselves are spherically symmetric and do not generate
gravitational waves but the collision events do. 

In addition, the Reynolds number of the cosmic plasma is very high and the
bubbles are expected to induce turbulence and, if there are small magnetic seed fields these
are amplified by turbulence leading to an MHD turbulent plasma with turbulent kinetic energy
and magnetic fields in equipartition~\cite{MHDturb}.

The correlation scale is expected to be~\cite{Caprini:2007xq}  of the order of 
$k_*=\beta =(\De\eta_*)^{-1}\simeq 100/\eta_* \sim 10^{-3}$Hz, which is close 
to the frequency of the peak sensitivity for the space born gravitational 
wave antenna {LISA}, proposed for launch in 2018 by ESA~\cite{LISA}.

\subsection{Bubble collisions}
\begin{figure}[ht]
\begin{center}
\includegraphics[width=14cm]{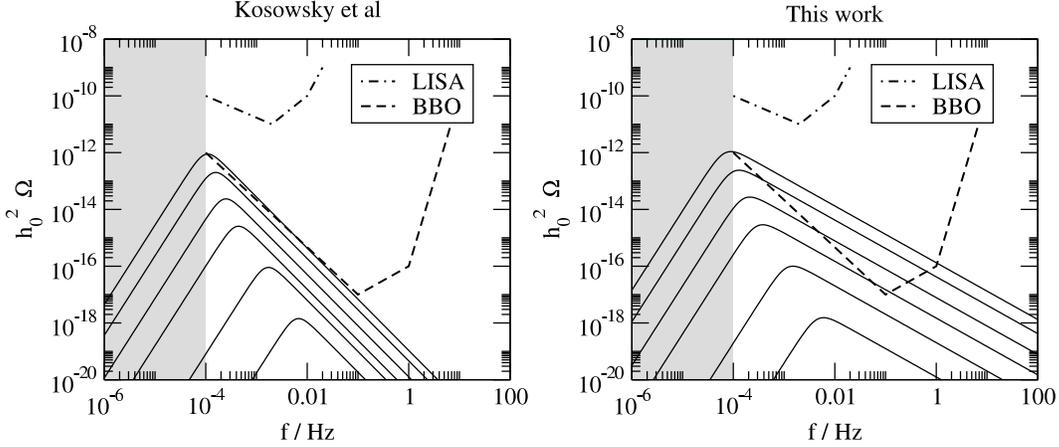} \\
\end{center}
\caption{\label{f:bubbles} $\frac{d\Om_{gw}(f)}{d\log f}$ from colliding bubbles, $f=k/2\pi$,
numerical results for $\Om_X/\Om_{\rm rad}=0.03$. On the left panel the old
results from Ref.~\cite{KKTW} are shown while the right panel shows the result of
Ref.~\cite{Huber:2008hg}. Figure from~\cite{Huber:2008hg}.}
\end{figure}

New numerical simulations of gravitational waves from bubble collisions during 
the electroweak phase transition~\cite{Huber:2008hg} have given a slower 
decay law on small scales, like $\propto k^{-1}$, see Fig.~\ref{f:bubbles}.

In Ref.~\cite{thomas} it has been shown that the decays law 
$\frac{d\Om_{gw}(k)}{d\log k} \propto k^{-1}$ is generically expected 
on small scales (large $k$) 
if $P_\Pi(\eta,\eta,k)$ is continuous in time but not differentiable. This behavior 
seems reasonable for bubble collisions which start from vanishing overlap
continuously but with a finite slope.

\subsection{Turbulent MHD}
Since the Reynolds number $Re$,  of the cosmic plasma at $T\sim 100$GeV is 
very high, the rapidly expanding bubbles of the broken phase provoke the formation
of turbulence. Furthermore, in the broken phase the electromagnetic field  
does generically not vanish. Since the magnetic Reynolds number $R_m$ is 
even higher than the Reynolds number of the plasma~\cite{geraldine2}, and 
thus the Prandl number~\cite{MHDturb} $P_m =R_m/Re\gg 1$, the high 
conductivity rapidly damps the electric fields and we are left with a 
magnetic field in a turbulent plasma, {\em MHD turbulence}~\cite{MHDturb}.

To determine the anisotropic stress tensors from the turbulent flow and from 
the magnetic field which is the source of gravitational waves, we first discuss
the vorticity and magnetic field spectra.
Because both, the vorticity and the magnetic field are divergence free, 
causality requires that their spectra behave like $k^2$ for small $k$: 
\bea 
\langle v_i(\bk)v_j(\bk')\rangle &=& (2\pi)^3\de^3(\bk-\bk')
(\hat{k}_j\hat{k}_i -\de_{ij}) P_v(k)\,, \\
\langle B_i(\bk)B_j(\bk')\rangle &=& (2\pi)^3\de^3(\bk-\bk')
(\hat{k}_j\hat{k}_i -\de_{ij})P_B(k)  \,.
\eea 
The functions $(\hat{k}_j\hat{k}_i -\de_{ij})P_\bullet (k)$ must be analytic 
at $k=0$ because of causality (the correlation functions are functions with 
compact support, they vanish for separations larger than the Hubble scales, 
thus their Fourier transforms must be analytic)  hence $P_v(k)$ and 
$P_B(k) \propto k^2$ for small $k$, see~\cite{DC03} . Turbulence is 
steered during the phase transition and then decays freely with decay time 
of the order of $\De\eta_*$. This source is not truly long-lived, but since it 
decays only like a power law and not exponentially, it does still contribute 
significantly for $\eta>\eta_*+\De\eta_*$ and we expect a suppression factor 
$\ep$ in eq.~(\ref{e:Omgwsimp}) which
is larger than the naive factor $(\HH_*\De\eta_*)^2$.  

The behavior of the spectrum on scales smaller than the correlation scale 
$k>k_*$ is expected to be a  {\em Kolmogorov spectrum} for the vorticity 
field, $P_v \propto k^{-11/3}$ and an {\em Iroshnikov--Kraichnan 
spectrum}~\cite{Iroshnikov-Kraichnan} for the magnetic field, 
$P_B \propto k^{-7/2}$. For the induced GW spectrum this yields
$$
\frac{d\Om_{GW\bullet}(k,\eta_0)}{d\ln(k)} \simeq \ep\; \Om_{\rm rad}(\eta_0)
 \left(\frac{\Om_{\bullet}(\eta_*)}{\Om_{\rm rad}(\eta_*)}\right)^2\times
\left\{\begin{array}{cc} (k/k_*)^3  & \mbox{ for } k\ll k_*  \\ 
(k/k_*)^{-\al}  & \mbox{ for } k\gg k_* \,.   \\  
\end{array} \right.
$$
For $\bullet =v$ we have $\al=11/3-1=8/3$ and for $\bullet =B$ we 
have $\al=7/2-1=5/2$, see~\cite{CDturb}. To find $\al$ we use that on small 
scale we expect the anisotropic stress spectrum to decay like the velocity, 
respectively magnetic field spectrum. With $\langle |\dot h|^2\rangle 
\propto P_\Pi/k^2$ and $\frac{d\Om_{gw}(k)}{d\ln(k)}\propto 
k^3\langle |\dot h|^2\rangle \propto kP_\Pi$ so that $\al =\ga-1$, where
$\ga$ denotes the decay exponent of the anisotropic stress spectrum.

\begin{figure}[ht]
\begin{center}
\includegraphics[width=13cm]{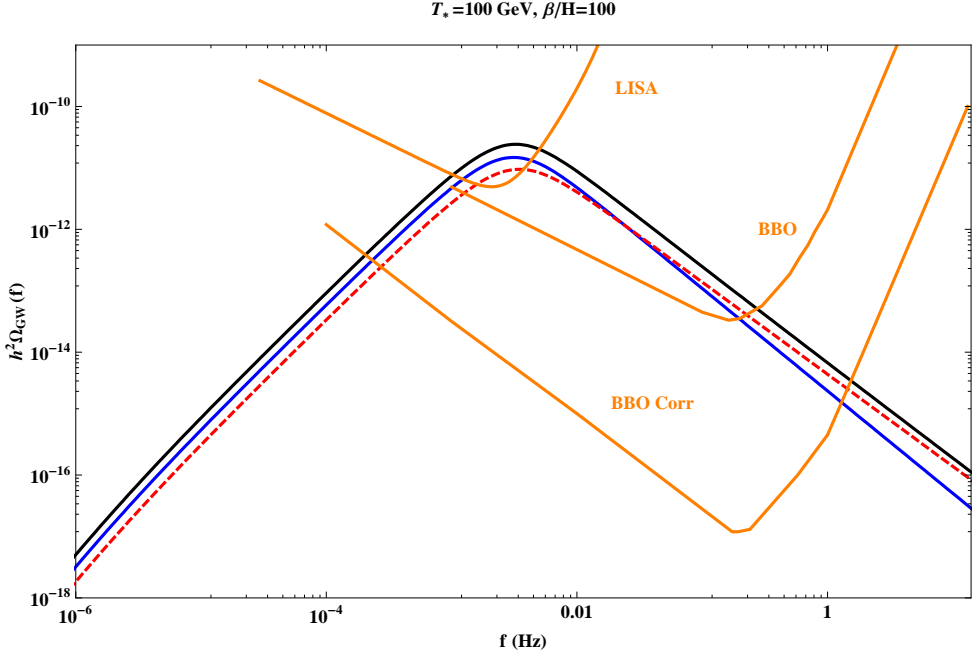} \\
\end{center}
\caption{\label{f:MHDturb}
 The gravitational wave energy density from magnetic fields (red) and 
turbulence (blue, the sum is shown in  black) generated during the 
electroweak phase transition at $T_*=100$GeV. The time-decorrelation 
of the source is modeled by the finite
correlation approach (Kraichnan decorrelation). The source energy density 
spectrum is assumed to be maximal, $\Om_v/\Om_{\rm rad} =2/9$, 
see~\cite{geraldine2}. Equipartition in turbulent kinetic energy and 
magnetic field energy is assumed. The sensitivity curves are 
from~\cite{Buonanno:2003th}. Figure taken from~\cite{geraldine2}.}
\end{figure}

\begin{figure}
\begin{center}
      \includegraphics[width=10cm]{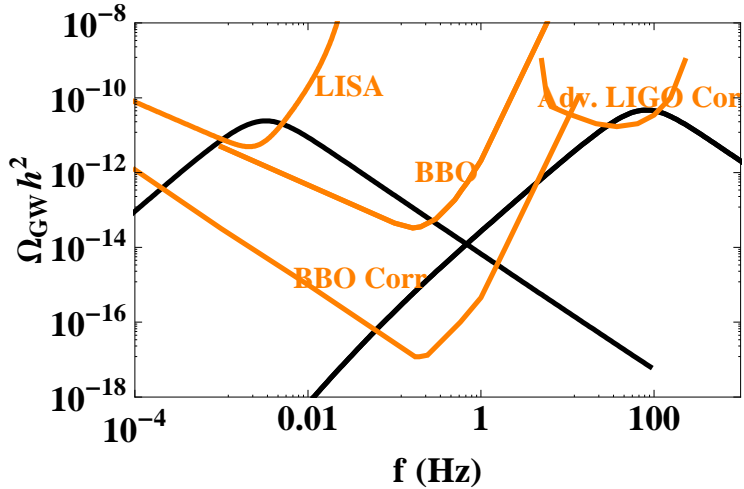} 
\end{center}
\caption{\label{f:MHD2} To the GW background from the electroweak phase 
transition also shown in Fig.~\ref{f:MHDturb} (here left black line) we add 
the result for a phase transition at $T=5 \times10^6$ GeV with 
$\De\eta_*\HH_*=0.02$ (black curve to the right). In this case, the 
gravitational wave background might even lie within the reach of advanced 
LIGO. The sensitivity curves are from~\cite{Buonanno:2003th}. 
}
\end{figure}

In Figs.~\ref{f:MHDturb} and \ref{f:MHD2} the gravitational spectrum 
obtained from the wlectroweak phase trnasityion and from a first order 
phase transition at $T_*=5 \times10^6$ GeV are shown and compared with future 
gravitational wave experiments like LISA~\cite{LISA}, the Bib Bang Observer 
(BBO)~\cite{BBO} asnd advances LIGO~\cite{LIGO}.

\subsection{Limits of primordial magnetic fields from the EW transition}
It is difficult to estimate $\Om_B(\eta_*)$ or $\Om_v(\eta_*)$ directly, but 
since causality requires the spectra to be so blue, the limit on gravitational 
waves which comes from small scales $k\simeq k_*$, 
yields very strong limits on primordial magnetic fields on large scales 
already from the simple {\em nucleosynthesis constraint}, 
$\Om_{gw} \lsim 0.1\Om_{\rm rad}$. If the magnetic field spectrum behaves 
like $P_B(k)\propto k^n$, the gravitational wave energy density per log interval 
must go like $\frac{d\Om_B(k,\eta_*)}{d\ln(k)} \propto k^{3+n}$ on large scales. 
For causal magnetic fields we have $n=2$, hence  the magnetic field energy 
spectrum is very blue, 
\be 
\frac{d\Om_B(k,\eta_*)}{d\ln(k)} \propto k^5 \quad \mbox{ for } \quad k<k_*\, .
\ee 
The main contribution to the gravitational wave energy density comes from the 
correlation scale, $\Om_{gw} \simeq \frac{d\Om_{gw}(k_*)}{d\ln(k_*)}$, whereas 
we are interested in the magnetic fields on much larger scales, for example
 $k_L =(0.1$Mpc$)^{-1}$. Using our simple formula~(\ref{e:Omgwsimp}) with 
$\ep\sim 1$ the nucleosynthesis limit yields 
$k_*^{3/2}B(k_*,\eta_0) \lsim 10^{-6}$Gauss. The time $\eta_0$ denotes today.
For the electroweak phase transition $k_* \simeq 100\HH_*\simeq 10^{-3}$sec$^{-1}$ 
while we are interested in magnetic fields on scales of $0.1$Mpc giving 
$k_L =10($Mpc$)^{-1} \simeq 10^{-13}$sec$^{-1}$.
With the above scaling, the limit on a magnetic field on this scale is
\be
k_L^{3/2}B(k_L,\eta_0) \lsim \left(\frac{k_L}{k_*}\right)^{5/2}10^{-6}{\rm Gauss} 
   =10^{-31}{\rm Gauss}\, . 
\ee  
Note also that this limit is very stable with respect to the small unknown 
pre-factor $\ep$. Since $\Om_{gw} \propto B^4$, changing $\ep$ by 2 orders 
of magnitude yields a change of only a factor of 3 in the limit for $B$.

This limit is unavoidable if the magnetic field on large scales changes
simply by flux conservation so that $B(\eta) =B(\eta_*)(a_*/a)^2$. Scaling 
this to today yields  $B(\eta_0) =B(\eta_*)/(1+z_*)^2$. 

\begin{figure}[ht]
\begin{center}
\includegraphics[width=9cm]{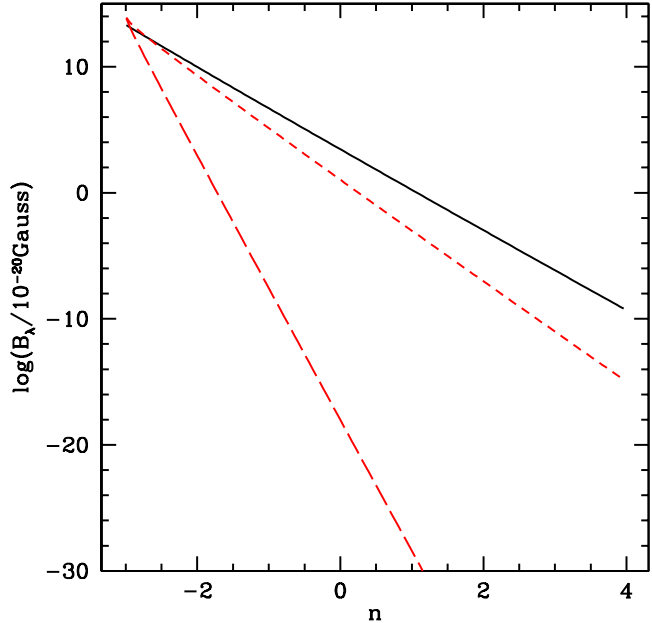}  
\end{center}
\caption{\label{f:lim}    The nucleosynthesis limit on magnetic fields from 
$\rho_B(t_{\rm nuc})<0.1\rho_{\rm rad}(t_{\rm nuc})$ (black solid line), 
and from the gravitational wave constraint (red lines) as a function of 
the spectral index $n$. The limits for magnetic fields from the electroweak 
phase transition are short-dashed (red), while the long dashed (red) 
line corresponds to magnetic fields from inflation at a GUT scale. 
Causality requires that the spectral index 
be $n=2$ for magnetic fields from the electroweak phase transition while 
$n$ is in principle undetermined for inflation. Figure from~\cite{GW1mag}.}
\end{figure}

The limit is so severe because the magnetic field spectrum is so red.
 The simple requirement that $\rho_B(t_*)<\rho_{\rm rad}(t_*)$ yields roughly 
the same limit. However, this limit is not entirely save since at very early 
time most of the energy density might in principle have been in the form of small 
scale magnetic fields which later are converted into heat by viscosity 
damping on small scales. The fact that such magnetic fields would generate a 
sizable background of gravitational waves, however, forbids this possibility.
This has been realized for the first time in Ref.~\cite{GW1mag}.
The inferred limits on magnetic fields as a function of the spectral index
are shown in Fig.~\ref{f:lim}.

\subsection{Helicity}
The most optimistic dynamo models for the generation of the magnetic fields of 
about $10^{-6}$Gauss which are observed in galaxies and clusters, need seed fields of 
the order of at least $10^{-21}$Gauss. These are excluded by the above limit.
There is a possible way out: If the magnetic field has non-vanishing helicity 
it can develop an inverse cascade and power can be transferred from
small to larger scales~\cite{BS}. In this case the large scale magnetic fields
do not simply decay by flux conservation in an expanding universe but they 
can feed from the small scale fields by inverse cascade.

During the electroweak phase transition parity is broken.
Actually, the Chern-Simons winding number of the gauge field, $N_{CS} \propto 
\int F\wedge A$, which is related to the baryon number, has an electromagnetic 
part to it which is simply the helicity of the magnetic 
field~\cite{vach}, $H = V^{-1}\int_V (A\cdot B) d^3x$ . It is interesting to note
that this relates the baryon number and the magnetic helicity.
It has been shown that the parity violation in the stress energy tensor 
coming from such helical magnetic fields lead to $T$-$B$ and $E$-$B$ 
correlations in the temperature and polarization spectra of the cosmic 
microwave background~\cite{tina}. By the same token they also 
generate gravitational waves with non-vanishing helicity~\cite{tina}.

\begin{figure}
\begin{center}
\includegraphics[width=7.5cm]{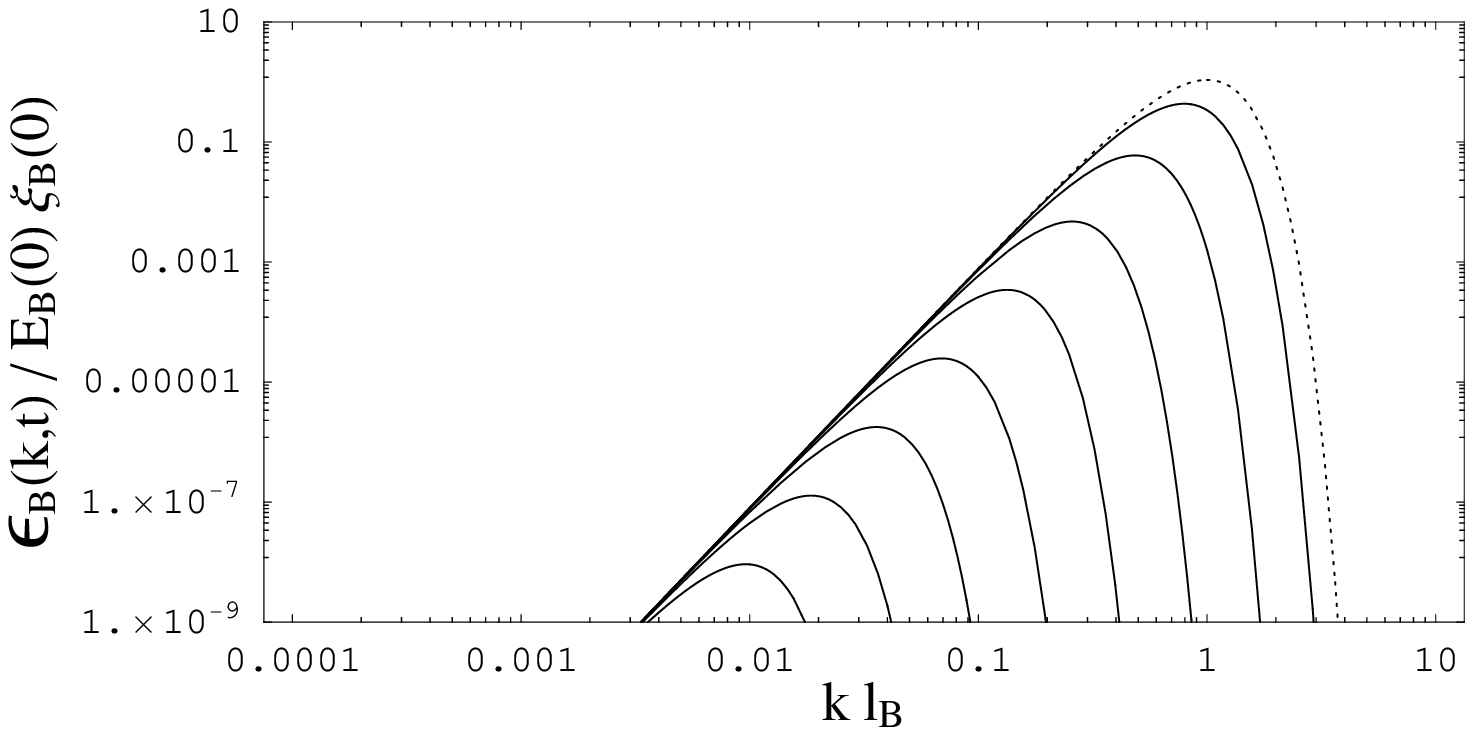}  \quad
\includegraphics[width=7.5cm]{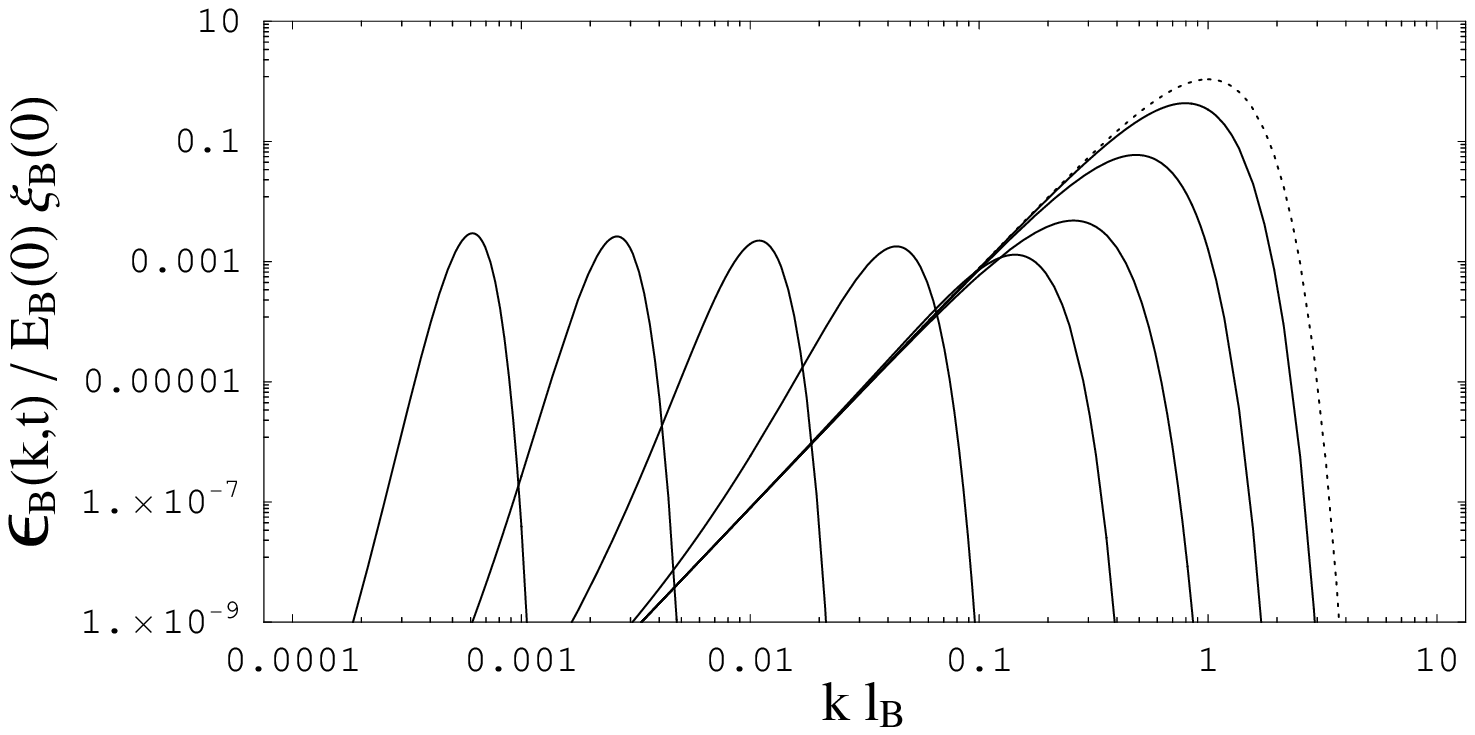} 
\end{center}
\caption{\label{f:camp}Left: the MHD evolution of the large scale magnetic 
field energy density spectrum with vanishing helicity. The field is simply damped 
by viscosity on small scales. Large scales are not effected. The viscosity
scale is growing with time.\\ 
Right: the case of non-vanishing helicity. First, the helicity is relatively 
small and the evolution of the helical field is not very different from the 
non-helical case. But due to helicity conservation, helicity rapidly becomes 
maximal and the evolution enters the inverse cascade regime. \\ For simplicity, 
the correlation scale is set equal to the damping scale. Figure 
from~\cite{campanelli}. }
\end{figure}

The inverse cascade which is caused by  helicity conservation for a helical 
field and is absent for non-helical magnetic fields has been investigated 
numerically, e.g., in~\cite{campanelli} and is shown in Fig.~\ref{f:camp}

In Ref.~\cite{campanelli} simple power laws are derived for the evolution
of the correlation scale in the inverse cascade regime.
Using this simple prescription, the mitigation of the magnetic field limits
by an inverse cascade of maximally helical magnetic fields has been 
estimated in Ref.~\cite{elisa}. It has been found, however, that the inverse 
cascade is not sufficient to present a way out for the magnetic fields from
the electroweak phase transition, but it can work for helical magnetic 
fields generated during the QCD phase transition. The QCD phase transition 
might be first order if the neutrino chemical potential is large, as
discussed in~\cite{DS}.

If the magnetic field is helical, the GW background would not be parity 
symmetric. There would be more GW's of one helicity than of the 
other~\cite{tina}.

\section{Conclusions}
\begin{itemize}
\item The rapid expansion and collision of bubbles during a first order phase transition stirs 
the relativistic cosmic plasma sufficiently to lead to the
generation of a (possibly observable) stochastic gravitational wave background. \vspace{0.1cm}
\item Observing such a background would open a new window to the early Universe 
and to high energy physics.  \vspace{0.1cm}
\item Generically, the density parameter of the GW background is of the order of
$$ \Om_{gw}(t_0) \simeq \ep \; \Om_{\rm rad}(t_0)
 \left(\frac{\Om_X(t_*)}{ \Om_{\rm rad}(t_*)}\right)^2\, ,$$
where $\ep<1$ is determined by the duration of the source, $\ep \simeq 1$ if
the decay time of the source is roughly a Hubble time.
 \vspace{0.1cm}
\item The spectrum grows like $\frac{d\Om_{gw}(k,t_0)}{d\ln(k)} \propto k^3$ on large scales
and decays on scales smaller than the correlations scale $k_* \sim 1/\De\eta_*$. The decay law 
depends on the physics of the source.
\item Within the standard model, the electroweak phase transition is not of 
first order and does (probably) not generate an appreciable gravitational 
wave background. However, simple deviations from the standard model can make 
it first order.  \\
In this case we expect a GW background which can marginally be 
detected by the LISA satellite. \vspace{0.1cm}
\item It has been proposed that the magnetic fields generated 
during the electroweak phase transition, 
could represent the seeds for the fields observed in galaxies and clusters. 
Unfortunately this idea does not withstand detailed scrutiny. The magnetic 
fields cannot have enough power on large scale to represent such 
seed fields. On small scales the magnetic field power which is initially 
present is dissipated later on by the viscosity of the cosmic plasma.
 \vspace{0.1cm}
\item If there is no inverse cascade acting on the magnetic field spectrum, 
the limits on the large scale fields coming from the generated GW background 
are too strong to allow significant magnetic fields even for the most 
optimistic dynamo mechanism.  \vspace{0.1cm}
\item If the magnetic field is helical, helicity conservation 
provokes an inverse cascade which can alleviate these limits.  Most probably
the relaxed limits are still too stringent for magnetic fields from the 
electroweak phase transition, but magnetic fields from the QCD
phase transition might be sufficient to seed the fields in galaxies and clusters
if they are helical. \vspace{0.1cm}
\item Helical magnetic fields generate a parity violating gravitational wave 
background, $|h_+(k)|^2 \neq |h_-(k)|^2$. This parity violation is in 
principle observable.
\end{itemize}

\ack I am grateful to my collaborators Chiara Caprini, Elisa Fenu, 
Tina Kahniashvili, Thomas Konstandin and Geraldine Servant who 
worked with me on most of the results reported here. I tremendously 
enjoyed the many hours of animated and stimulating discussions 
which helped us to gain insight in this subject which touches on so many 
fascinating areas of physics. I thank Leonardo Campanelli, Chiara Caprini 
and Geraldine Servant for helpful comments on a first draft of this paper.
Finally, it is a pleasure to thank the organizers of the First Mediterranean 
Conference on 
Classical and Quantum Gravity (MCCQG) for inviting me to give this talk 
and to participate in a lively conference on beautiful Crete. 
My work is supported by the Swiss National Science Foundation.

\end{document}